\documentclass[a4paper,12pt]{article}
\usepackage[margin=1in]{geometry}
\usepackage{graphicx}

\usepackage{amssymb}
\usepackage{color}
\usepackage{multirow}
\usepackage{amsmath}
\usepackage{breqn}
\usepackage{bm}
\usepackage{ascmac}
\usepackage[colorlinks=true, urlcolor=blue]{hyperref}
\usepackage{wrapfig}
\usepackage{url}
\usepackage{color}
\usepackage[modulo]{lineno}
\usepackage{threeparttable}
\usepackage{float}
\usepackage{siunitx}
\usepackage{subfig}
\usepackage{here}
\usepackage{comment}
\usepackage{authblk}

\newcommand{\klpionn}{K_L \to \pi^0 \nu \overline{\nu}}

\newcommand{\klpioee}{K_L \to \pi^0 e^+ e^-}

\newcommand{\kpluspnn}{K^+ \to \pi^+ \nu \overline{\nu}}

\title{Experimental Study of Rare Kaon Decays at J-PARC with KOTO and KOTO~II}

\author[1]{J.~K.~Ahn}
\author[2]{E.~Augustine}
\author[3]{L.~Bandiera}
\author[4]{J.~Bian}
\author[5]{F.~Brizioli}
\author[3]{N.~Canale}
\author[6]{G.~A.~Carini}
\author[7]{V.~Chobanova}
\author[8]{G.~D'Ambrosio}
\author[9,10]{J.~B.~Dainton}
\author[11]{S.~De~Capua}
\author[3,12]{P.~Fedeli}
\author[3]{A.~Gianoli}
\author[13]{A.~Glazov}
\author[14]{M.~Gonzalez}
\author[15]{E.~Goudzovski}
\author[14]{M.~Homma}
\author[16]{Y.~B.~Hsiung}
\author[15,17]{T.~Husek}
\author[18]{A.~M.~Iyer}
\author[19]{E.~J.~Kim}
\author[20]{C.~Kim}
\author[21,22]{T.~K.~Komatsubara}
\author[14]{K.~Kotera}
\author[23]{M.~Kreps}
\author[15]{C.~Lazzeroni}
\author[21]{G.~Y.~Lim}
\author[20]{S.~Lim}
\author[24]{C.~Lin}
\author[25,26]{F.~Mahmoudi}
\author[3]{L.~Malagutti}
\author[7]{D.~Martinez~Santos}
\author[9]{K.~Massri}
\author[27]{T.Matsumura}
\author[3,12]{A.~Mazzolari}
\author[28]{M.~Moulson}
\author[14]{H.~Nanjo}
\author[3,12]{R.~Negrello}
\author[25]{S.~Neshatpour}
\author[21,22]{T.~Nomura}
\author[14]{D.~Ogawa}
\author[14]{K.Ono}
\author[3]{G.~Paternò}
\author[5]{M.~Pepe}
\author[29]{L.~Peruzzo}
\author[30]{D.~Protopopescu}
\author[2]{J.~Redeker}
\author[3,12]{M.~Romagnoni}
\author[15]{A.~Romano}
\author[13]{A.~Rostomyan}
\author[9]{A.~Shaikhiev}
\author[21,22]{K.~Shiomi}
\author[14]{R.~Shiraishi}
\author[28]{M.~Soldani}
\author[2]{B.~Stillwell}
\author[28]{J.~Swallow}
\author[3]{A.~Sytov}
\author[31]{Y.~Tajima}
\author[15]{A.~Tomczak}
\author[32]{Y.~C.~Tung}
\author[2]{Y.~W.~Wah}
\author[29]{R.~Wanke}
\author[21]{H.~Watanabe}
\author[14]{T.~Yamanaka\thanks{Present address : International Affairs Division, High Energy Accelerator Research Organization (KEK), Tsukuba, Ibaraki 305-0801, Japan}}
\author[6]{G.~Yang}
\affil[1]{Department of Physics, Korea University, Anam 145, Seongbuk-gu, Seoul 02841, Republic of Korea}
\affil[2]{Enrico Fermi Institute, University of Chicago, 5640 S Ellis Ave, Chicago, IL 60637 USA}
\affil[3]{INFN Sezione di Ferrara, Via Saragat 1, 44122 Ferrara, Italy}
\affil[4]{University of California, Irvine, CA 92697-4575, USA}
\affil[5]{INFN Sezione di Perugia, 06123 Perugia, Italy}
\affil[6]{Brookhaven National Laboratory, Instrumentation Department, 20 Technology St, Upton NY 11973, USA}
\affil[7]{Universidade da Coru\~{n}a, Campus Industrial, Departamento de F\'isica y Ciencias de la Tierra, Ferrol 15471, Spain}
\affil[8]{INFN Sezione di Napoli, Italy}
\affil[9]{Department of Physics, Lancaster University, Lancaster, LA1 4YW, United Kingdom}
\affil[10]{Cockcroft Institute, UKRI Daresbury Laboratory, Warrington WA4 4AD, United Kingdom}
\affil[11]{Department of Physics and Astronomy, The University of Manchester, Manchester, M13~9PL, United Kingdom}
\affil[12]{Department of Physics and Earth Science, University of Ferrara, Via Saragat 1, Ferrara, 44122, Italy}
\affil[13]{Deutsches Elektronen-Synchrotron DESY, Notkestr. 85, 22607 Hamburg, Germany}
\affil[14]{Department of Physics, Osaka University, Toyonaka, Osaka 560-0043, Japan}
\affil[15]{School of Physics and Astronomy, University of Birmingham, Edgbaston, Birmingham B15~2TT, United Kingdom}
\affil[16]{Department of Physics, National Taiwan University,  Taipei, 10617 Taiwan}
\affil[17]{Institute of Particle and Nuclear Physics, Charles University, V Hole\v{s}ovi\v{c}k\'ach 2, 180 00 Prague, Czech Republic}
\affil[18]{Department of Physics, Indian Institute of Technology Delhi, Hauz Khas, New Delhi-110016, Delhi, India}
\affil[19]{Division of Science Education, Jeonbuk National University, Jeonju 54896, Republic of Korea}
\affil[20]{Department of Physics, Pusan National University, Busan 46241, Republic of Korea}
\affil[21]{Institute of Particle and Nuclear Studies, High Energy Accelerator Research Organization (KEK), Tsukuba, Ibaraki 305-0801, Japan}
\affil[22]{J-PARC Center, Tokai, Ibaraki 319-1195, Japan}
\affil[23]{Department of Physics, University of Warwick, Coventry, CV4 7AL, United Kingdom}
\affil[24]{Department of Physics, National Changhua University of Education, Changhua 50007, Taiwan}
\affil[25]{Universit\'e Claude Bernard Lyon 1, CNRS/IN2P3, Institut de Physique des 2 Infinis de Lyon, UMR 5822, F-69622, Villeurbanne, France}
\affil[26]{Theoretical Physics Department, CERN, CH-1211 Geneva 23, Switzerland}
\affil[27]{Department of Applied Physics, National Defense Academy of Japan, 1-10-20 Hashirimizu, Yokosuka, Kanagawa 238-0006}
\affil[28]{INFN Laboratori Nazionali di Frascati, 00044 Frascati RM, Italy}
\affil[29]{Institute of Physics and PRISMA+ Cluster of Excellence, University of Mainz, 55099 Mainz, Germany}
\affil[30]{School of Physics and Astronomy, University of Glasgow, Glasgow G12 8QQ, United Kingdom}
\affil[31]{Faculty of Science, Yamagata University, 1-4-12 Kojirakawa,Yamagata 990-8560, Japan}
\affil[32]{Department of Physics, National Kaohsiung Normal University, Kaohsiung 824, Taiwan}

\begin{document}
\maketitle
\begin{abstract}
The rare kaon decay $K_L\to\pi^0\nu\bar{\nu}$ is extremely sensitive to new physics, because the contribution to this decay in the Standard Model (SM) is highly suppressed and known very accurately; the branching ratio is $3\times 10^{-11}$ in the SM with a theoretical uncertainty of just 2\%.
The measurement of this branching ratio could provide essential new information about the flavor structure of the quark sector
from the $s\to d$ transition.
The decay is being searched for in the KOTO experiment at J-PARC, which has obtained the current best upper limit on the branching ratio of $2.2\times 10^{-9}$; a sensitivity to branching ratios below $10^{-10}$ is achievable by the end of the decade.
A next-generation experiment at J-PARC, KOTO~II, was proposed in 2024 with 82 members worldwide, including significant contributions from European members.
The goal of KOTO~II is to measure the $K_L\to\pi^0\nu\bar{\nu}$ branching ratio with sensitivity below $10^{-12}$ in the 2030s.
Discovery of the decay with $5\sigma$ significance is achievable at the SM value of the branching ratio.
An indication of new physics with a significance of 90\% is possible if the observed branching ratio differs by 40\% from the SM value.
Another important goal of KOTO~II is to measure the branching ratio of the unobserved $K_L\to \pi^0e^+e^-$ decay,
which can give an input to flavor structures of new physics. Other rare $K_L$ decays and hidden-sector particles are also in the scope of the study.
After 2026, KOTO will be the only dedicated rare kaon decay experiment in the world, and KOTO~II is the only future rare kaon decay project currently proposed. We would like to lead a global initiative for the experimental study of rare kaon decays, with significant contributions and support from the European community.
\end{abstract}
\section{Introduction}
The rare kaon decay $K_L\to\pi^0\nu\bar{\nu}$ is a sensitive probe for the exploration of physics beyond the Standard Model (SM) and is one of the key inputs to flavor physics. 
The decay is being searched for with the KOTO experiment~\cite{KOTOproposal} at J-PARC, Japan.
The current best upper limit on the branching ratio for this decay is $2.2\times 10^{-9}$, obtained with KOTO data taken in 2021~\cite{KOTO:2024zbl}. 
KOTO will achieve a sensitivity to branching ratios below $10^{-10}$ by the end of the decade, while the branching ratio predicted by the SM is $3\times 10^{-11}$. 
A next-generation experiment, KOTO~II~\cite{KOTO:2025gvq}, was proposed in 2024 
with the goal of observing the decay with more than $5\sigma$ significance in the 2030s at the SM value of the branching ratio.
The KOTO~II proposal was prepared in 2024, with 82 members from 11 countries worldwide, including significant contributions from European members.
The observation of the decay $K_L\to\pi^0e^+e^-$ is another important aim of the KOTO~II experiment, together with searches for other rare decays and hidden-sector particles.
After 2026, the KOTO experiment will be the only running experiment dedicated to rare kaon decays,
and the KOTO~II experiment is currently the only planned future kaon experiment in the world.
The KOTO-II Collaboration 
is intended as a global effort in kaon physics with significant contributions and support from the European community.

\section{Physics motivation}
The origin of the flavor structure (mass, mixing, and CP violation) in the SM is one of the current mysteries in particle physics, together with the large baryon asymmetry in the universe.
Inputs from $s\to d$ flavor transitions are indispensable to shed light on the flavor structure of the quark sector. 
Experimental studies of kaon decays have been and will continue to be important to provide these inputs. 
The $s\to d$ transition is a flavor changing neutral current. Due to GIM suppression, the dominant contribution is via higher-order Feynman diagrams with $s\to t\to d$ transitions; it is further suppressed because of the smallness of the corresponding CKM matrix elements. Because the SM contribution is highly suppressed and accurately known, rare kaon decays are particularly sensitive to contributions from new physics, therefore providing essential new insight into the flavor structure of the quark sector from $s\to d$ transition. 

The decay $K_L\to\pi^0\nu\bar{\nu}$ is 
the primary subject of KOTO and KOTO~II.
This decay is
a golden mode among other rare kaon decays because of the highly suppressed and accurately known SM contribution.
The $K_L$ is approximately $\left(K^0 - \bar{K}^0\right)/\sqrt{2}$, and the $K_L\to\pi^0\nu\bar{\nu}$ decay occurs due to interference between the two amplitudes from $K^0$ and $\bar{K}^0$. This decay is CP violating in the SM, with the only contributions from diagrams with an intermediate top quark.
In the SM, the branching ratio is $3\times 10^{-11}$ with a theoretical uncertainty of 2\%
~\cite{Anzivino:2023bhp,Buras:2022qip}
\footnote{$(2.94\pm 0.15)\times 10^{-11}$ from \cite{Buras:2022qip}. $2.87(7)(2)(23) \times 10^{-11}$ or $2.78(6)(2)(29) \times 10^{-11}$ from \cite{Anzivino:2023bhp} depending on the CKM parameter fits, where the values in parentheses are the short distance, long distance, and parametric uncertainties.}.
In new-physics scenarios in which the flavor transition $s\to d$ occurs at lower order at high mass scales, rare kaon decays are sensitive to new physics at energy scales as high as 1000~TeV. 
If, instead, the new-physics contribution to this flavor transition is suppressed in the same way as the SM contribution due to a common flavor symmetry, the correlations with other flavor transitions are key for the exploration of new physics. 

The importance of the decay $K^+\to \pi^+\nu\bar{\nu}$ is complementary, with both CP-violating and CP-conserving contributions and a small contribution from charm diagrams.
The branching ratio is $8\times 10^{-11}$ in the SM~\cite{Anzivino:2023bhp}.
This decay has been observed at $5\sigma$ significance
by the NA62 experiment at CERN with data collected from 2016 to 2022 to obtain a measured value of the branching ratio of
$13.0^{+3.3}_{-3.0}\times 10^{-11}$~\cite{NA62:2024pjp}.
NA62 data taking will continue until mid-2026, and a precision on the $K^+\to \pi^+\nu\bar{\nu}$ branching ratio of 15\% is expected to be reached.
A correlation between the charged and neutral $K\to\pi\nu\bar{\nu}$ branching ratios exists due to isospin symmetry. This indirectly limits $\mathcal{B}(K_L\to\pi^0\nu\bar{\nu})$ 
to be less than $4.3\times \mathcal{B}(K^+\to \pi^+\nu\bar{\nu})$, which is known as the ``Grossman-Nir bound''~\cite{ref:Grossman}. The combined values of the branching ratios for the charged and neutral decays can help to distinguish between different new-physics models if one or both values is found to be different from its SM prediction.

The decays $K_L\to\pi^0 e^+e^-$ and $K_L\to\pi^0\mu^+\mu^-$ are also sensitive probes of new physics, sharing the same or similar short distance contributions, and with both CP-violating and CP-conserving contributions. These will also be studied at KOTO~II, and are the main subject of the second phase.
The branching ratios are $\mathcal{O}(10^{-11})$ in the SM~\cite{Anzivino:2023bhp}.
A long-distance contribution with a virtual photon decaying to lepton pairs exists; its amplitude can be studied with the decay $K_S\to\pi^0\mu^+\mu^-$.

\section{New physics contributions}
\begin{figure}[tbph]
 \begin{center}
  \includegraphics[width=0.5\textwidth]{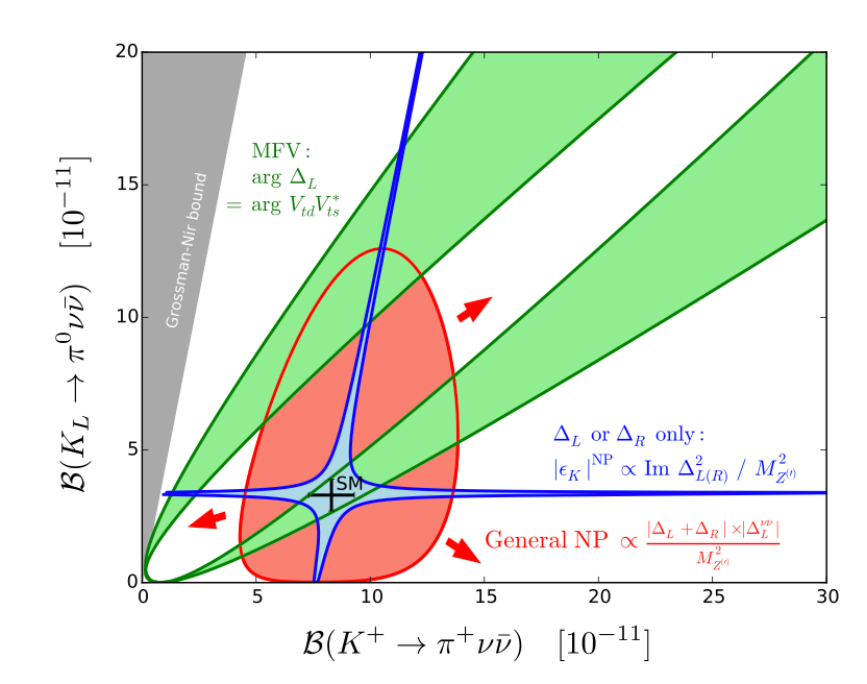}
  \end{center}
    \caption{Correlation between $\mathcal{B}(\klpionn)$ and $\mathcal{B}(\kpluspnn)$ for various new physics models.
    The blue region shows the correlation coming from the constraint by the $K$-$\overline{K}$ mixing parameter $\epsilon_{K}$ if
    only left-handed or right-handed couplings are present. The green region shows the correlation for models having
    a CKM-like structure of flavor interactions.  The red region shows the lack of correlation for models with general left-handed and
     right-handed couplings~\cite{Buras:2015yca}. }
  \label{fig:newphysics}
\end{figure}

\begin{figure}[h]
\centering 
\includegraphics[width=0.5\textwidth]{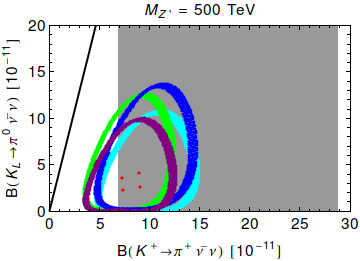}
\caption{Possible contributions from new physics in a Z' model are shown with four color bands corresponding to different sets of CKM parameters~\cite{Buras:2014zga}.\label{fig:500TeV}}
\end{figure}

\begin{figure}[h]
\centering 
\includegraphics[width=0.5\textwidth]{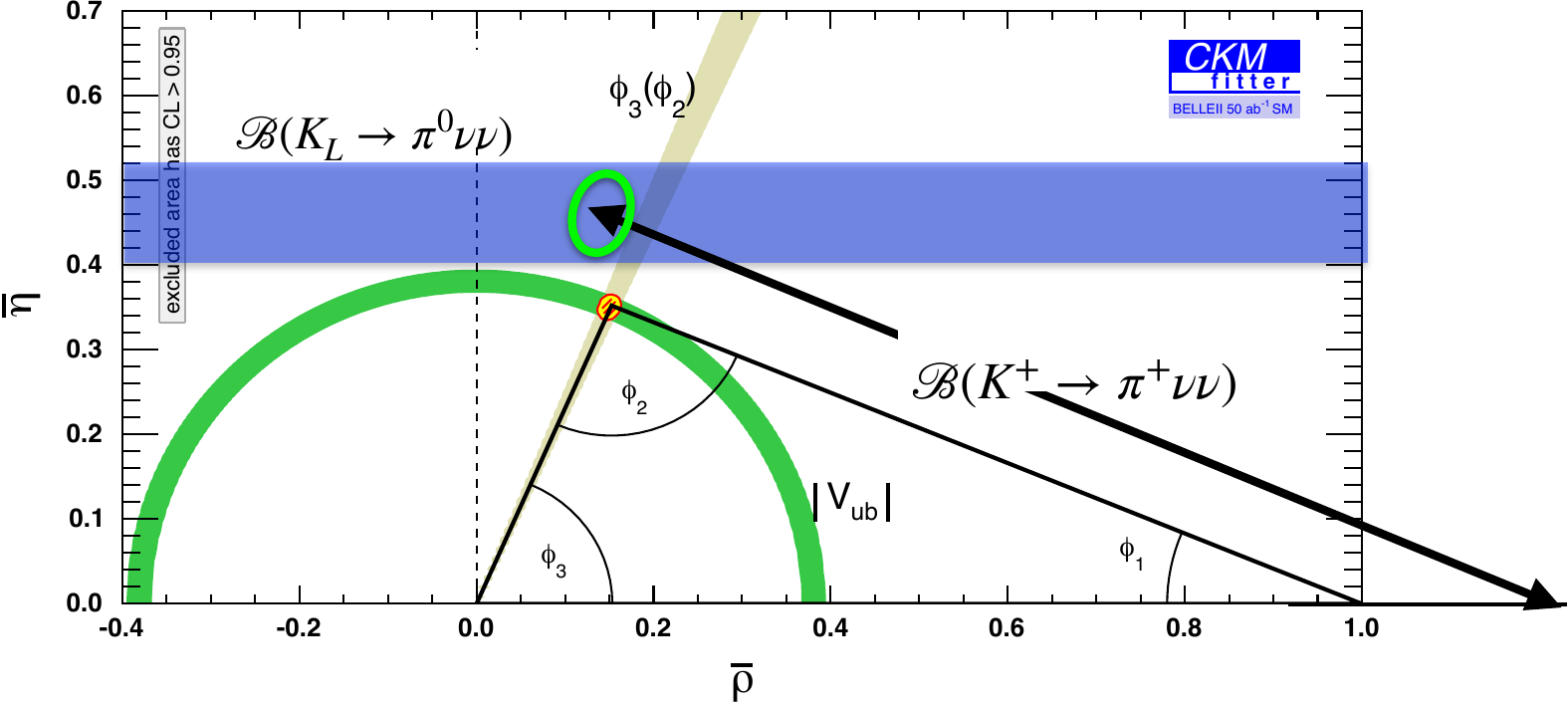}
\caption{Projection of the  unitarity triangle with a possible Belle~II input with tree diagrams~\cite{Belle-II:2018jsg}.
A concept of the unitarity triangle from kaon is illustrated  by overplaying possible constraints from the rare kaon decays $\klpionn$ and $\kpluspnn$.\label{fig:unitarityTriangle}
}
\end{figure}

\begin{figure}[h]
\centering 
\includegraphics[width=0.5\textwidth]{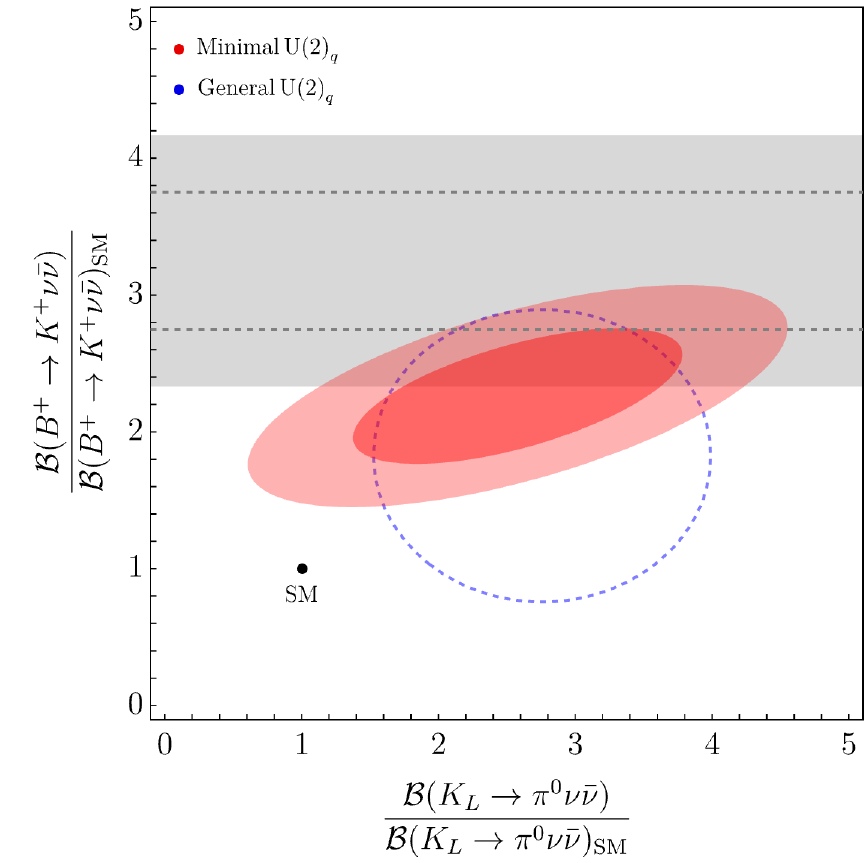}
\caption{
Correlation between $B^+\to K^+\nu\bar\nu$ and $K_L\to\pi^0\nu\bar\nu$ decay rates (normalized to their SM expected values) in several NP scenarios~\cite{Allwicher:2024ncl}. The red areas denote the parameter regions favored at $1\sigma$ and $2\sigma$ from a global fit in the limit of minimal $U(2)_q$ breaking. Dashed and dotted blue curves are $1\sigma$ and $2\sigma$ regions from a global fit where breaking is a free parameter. The gray bands indicate the current experimental constraints.}
\label{fig:newPhysBK}
\end{figure}

The $s\to d$ flavor transition can be studied with kaon decays to obtain insight on the possible differences between the SM and new-physics contributions.
Among rare kaon decays, $K_L\to\pi^0\nu\bar{\nu}$ is 
of particular importance, because it is purely CP violating and the SM value of its branching ratio can be calculated very precisely.
\begin{itemize}
\item The correlation between the branching ratios of $K_L\to\pi^0\nu\bar{\nu}$ and $K^+\to\pi^+\nu\bar{\nu}$~\cite{Buras:2015yca} (Fig.~\ref{fig:newphysics}) are different for different models of new physics, so measurements of both branching ratios may help to discriminate between new physics scenarios.  Physics beyond the SM (BSM) involving
SUSY~\cite{Endo:2017ums,Crivellin:2017gks,Tanimoto:2016yfy},
2HDM~\cite{Chen:2018ytc}, leptoquark~\cite{Fajfer:2018bfj,He:2018uey},
or right-hand neutrino~\cite{He:2018uey} could contribute.
\item Possible contributions from BSM physics at an energy scale of 1000~TeV can be probed~\cite{Buras:2014zga} (Fig.~\ref{fig:500TeV}), as is the case for new physics contributing to the decay at tree level and with a flavor transition coupling of $\mathcal{O}(1)$.
\item A unitarity triangle can be constructed with the branching ratios of the two $K$ decays, and can be compared to that from $B$ decays, revealing new physics contributions (Fig.~\ref{fig:unitarityTriangle}).
\item  In some new physics models with an approximate global flavor symmetry, the predicted values of the branching ratios for the $K_L\to \pi^0\nu\bar{\nu}$, $K^+\to \pi^+\nu\bar{\nu}$, and $B^+\to K^+\nu\bar{\nu}$ decays are correlated~\cite{Allwicher:2024ncl}. Current data suggests that the latter two decays have measured branching ratios larger than predicted in the SM, which suggests a larger branching ratio for the $K_L\to\pi^0\nu\bar{\nu}$ decay as well (Fig.~\ref{fig:newPhysBK}).
\item The decay $K_L\to \pi^0\nu\bar{\nu}$ is sensitive to new sources of CP violation, which may give a hint to source of the baryon asymmetry of the universe. 
Some models of electroweak baryogenesis that explain the current baryon asymmetry of the universe give deviations on the branching ratio of this decay~\cite{Kanemura:2023juv}.
\end{itemize}

\begin{figure}[h]
\centering
\includegraphics[width=0.47\textwidth]
{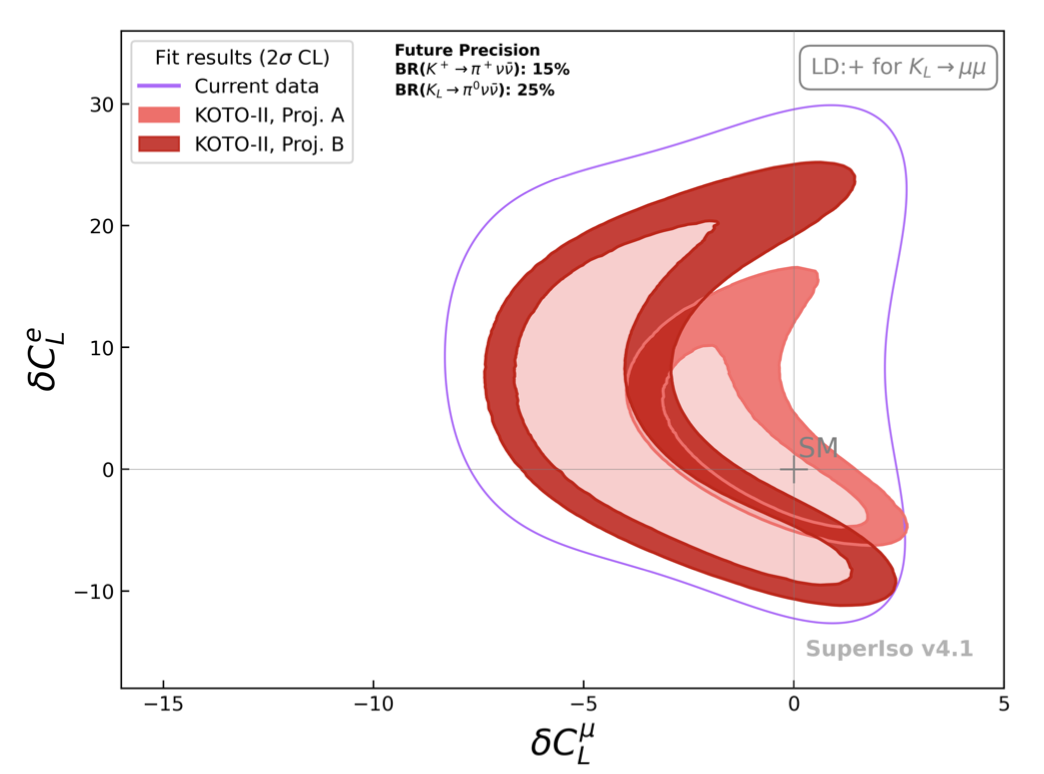}
\hspace*{3mm}
\includegraphics[width=0.47\textwidth]
{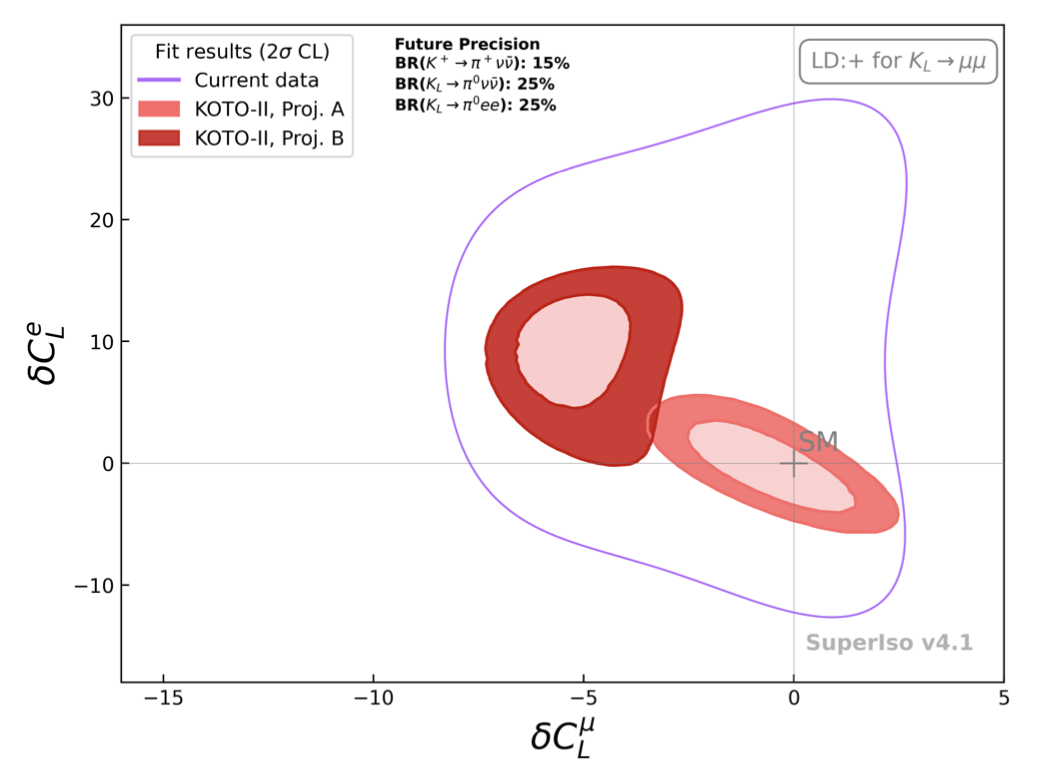} 
\caption{BSM parameter space for Wilson coefficients in scenarios with LFU violation where the NP effects for electrons are different from the those for muons and taus~\cite{DAmbrosio:2022kvb,DAmbrosio:2024ewg}. Left: Impact on allowed parameter space from measurements of the golden channels $K\to\pi\nu\bar\nu$ from NA62 and KOTO~II with the expected final precision. Right: Impact on the parameter space by the inclusion in the fits of a measurement of the $K_L\to\pi^0 e^+e^-$ branching ratio with 25\% precision.}
\label{fig:LFU}
\end{figure}

\paragraph{Sensitivity to other new physics scenarios.}
The rare decay $K_L\to\pi^0\nu\bar{\nu}$ is also sensitive to BSM physics in a variety of scenarios:
\begin{itemize}
\item 
lepton flavor non-universality through coupling differences to $\nu_e$, $\nu_\mu$, and $\nu_\tau$~\cite{DAmbrosio:2022kvb,DAmbrosio:2024ewg}
(Fig.~\ref{fig:LFU}),
\item 
lepton number violation~\cite{Deppisch:2020zrd},
\item 
invisible or long-lived neutral light bosons via the decay 
$K_L\to\pi^0 X^0$~\cite{Fuyuto:2014cya,Fuyuto:2015gmk,Kitahara:2019lws, Egana-Ugrinovic:2019wzj},
\item axions or axion-like particles with decays to two photons, since these particles can be produced similarly to $K_L$ in the experiment~\cite{Afik:2023mhj}.
\end{itemize}

\section{The KOTO and KOTO~II experiments}
\begin{figure}[h]
\centering
\includegraphics[width = 0.8\textwidth]{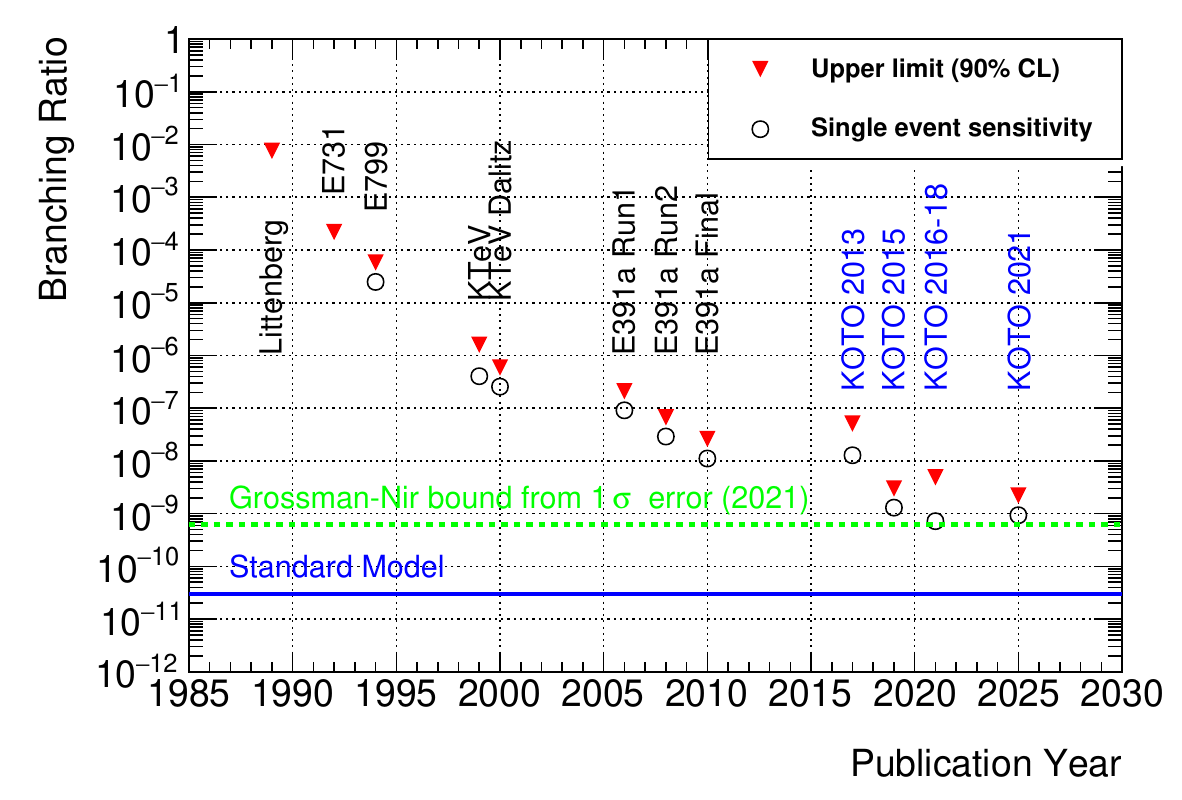}
\caption{History of the search for $\klpionn$.\label{fig:history}}
\end{figure}

\paragraph{The KOTO experiment}
Identification of the ultra-rare decay $\klpionn$ is difficult,
due to the enormous amount of background ($10^{10}$ times larger than the signal)
and limited kinematic information obtained (no initial state information with tracking,
missing neutrinos, and only two photons from a $\pi^0$ in the final state).
In order to overcome these difficulties, it is natural to envision a dedicated experiment.
Such dedicated experiments have been developed in Japan since the early 2000s (Fig.~\ref{fig:history}).

\begin{figure}[h]
\centering 
\includegraphics[width = 0.8\textwidth]{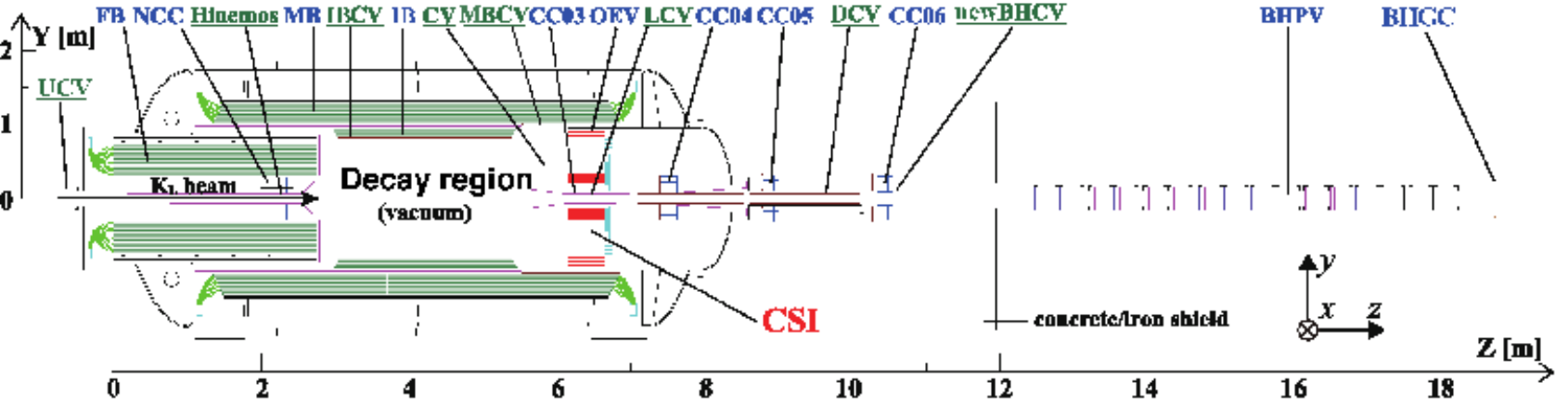}
\caption{Cross-sectional view of the KOTO detector.\label{fig:KOTOdetector}}
\end{figure}

\begin{figure}
    \centering
    \includegraphics[width=0.8\linewidth]{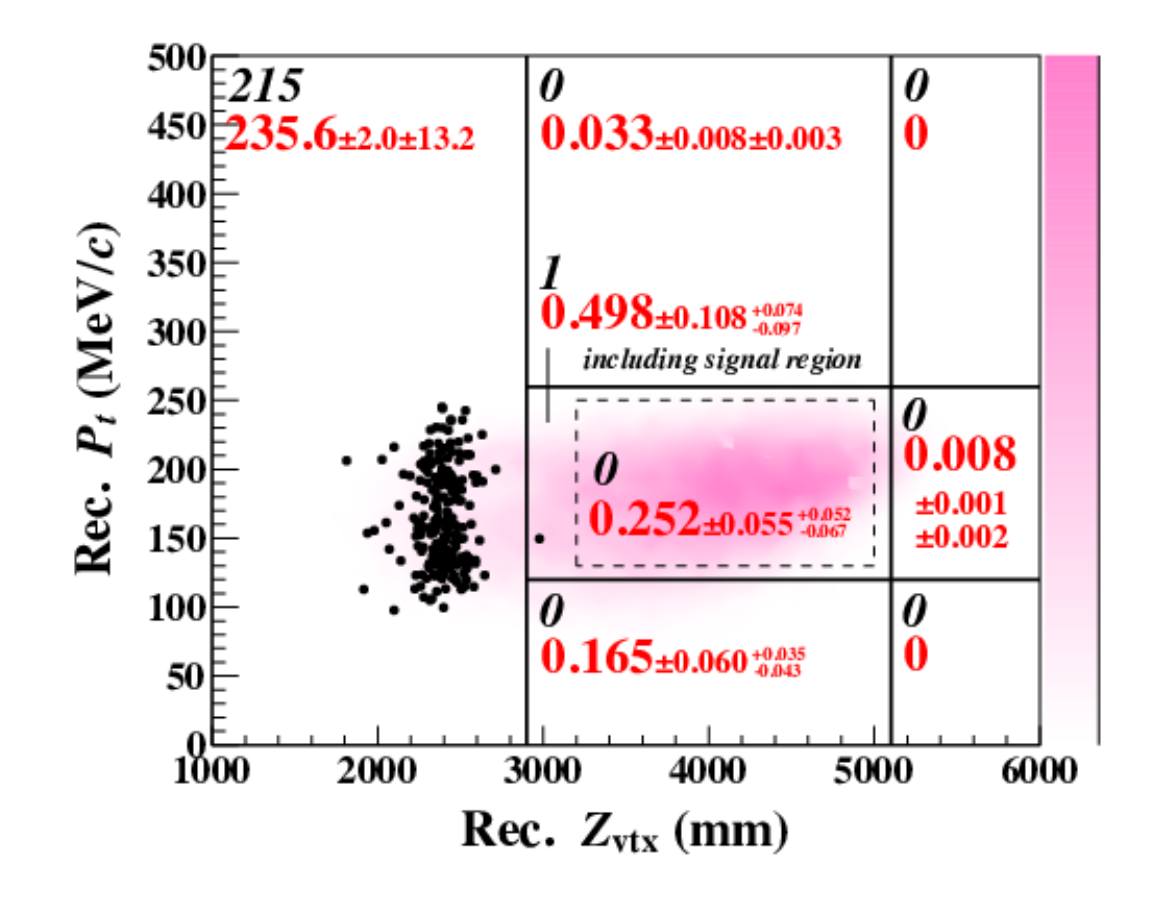}
    \caption{Distribution in the plane of reconstructed vertex position and transverse $\pi^0$ momentum for the analysis in the KOTO experiment with data takin in 2021.
    The region surrounded by the dashed lines is the signal region. The black dots represent the observed
events, and the shaded contour indicates the $\klpionn$ distribution from the MC simulation. The black italic (red regular) numbers indicate the number of observed (background) events for each region. The second and third red regular numbers indicate the statistical uncertainties and systematic uncertainties, respectively.
    }
    \label{fig:ptz}
\end{figure}

The KEK E391a experiment was the first such dedicated experiment. The KOTO experiment at J-PARC is its successor and is improving the sensitivity to the $\klpionn$ decay.
A neutral beam with photons, neutrons, and $K_L$s is extracted at 16 degrees from collisions of 30~GeV primary protons on a gold target.
Charged particles in the beamline are swept out with magnets,
while a two-stage collimator is used to define a narrow beam.  
In the KOTO detector as shown in Fig.~\ref{fig:KOTOdetector}, 
the two photons from a $\pi^0$ are detected with an electromagnetic calorimeter at the downstream of a 2-m long signal decay volume.
Backgrounds are reduced with a hermetic veto system consisting of counters surrounding the decay volume. The decay $\pi^0$ is reconstructed by assuming the vertex on the beam axis (z axis). Candidate events are defined in the plane of vertex position and $\pi^0$ transverse momentum (Fig.~\ref{fig:ptz}). Further background reduction is obtained with the kinematics of the $\pi^0$ and the cluster and pulse shapes in the calorimeter.

The experimental method was established with the E391a and KOTO experiments. 
A sensitivity to branching ratios below $10^{-10}$ is expected to be achieved by the end of the decade.
However, further progress will be limited due to the low signal statistics and background contributions.

\begin{figure}[h]
\centering
\includegraphics[width = 0.8\textwidth]{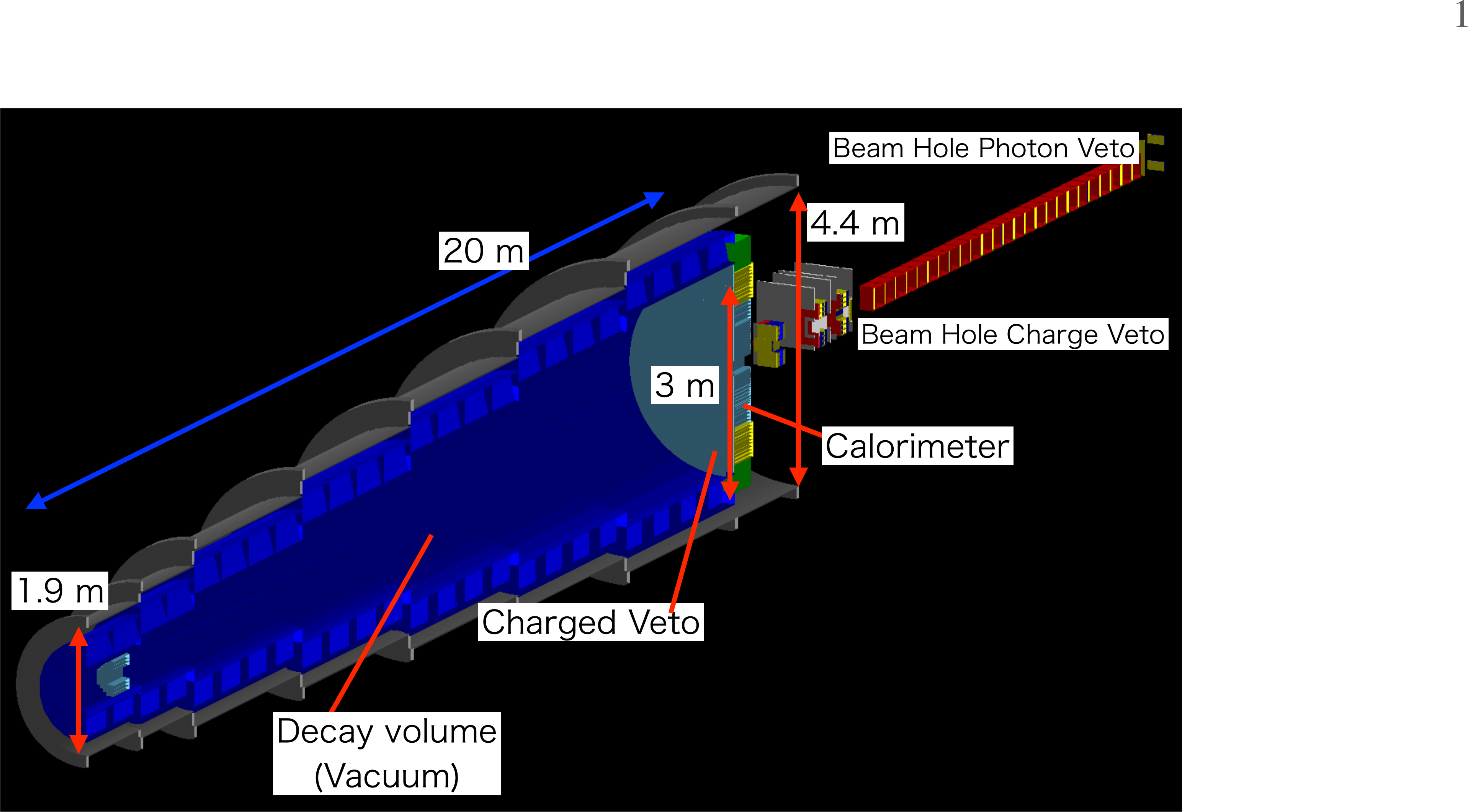}
\caption{The KOTO~II detector.\label{fig:kotoIIdet}}
\end{figure}

\paragraph{The KOTO~II experiment}
A next generation experiment at J-PARC, KOTO~II, was proposed in 2024 by a collaboration with 82 members 
from 11 countries worldwide,
with the goal of measuring the $\klpionn$ branching ratio at sensitivity below $10^{-12}$ during the 2030s.
The $K_L$ production angle is reduced to 5 degrees in the new beamline, which  gives 2.6 times more $K_L$ flux. It also gives a harder $K_L$ momentum spectrum peaking at 2.9~GeV/c, which allows the longer decay region to be used efficiently.
The KOTO~II detector is designed with a 12~m long signal decay region and 3~m diameter calorimeter (Fig.~\ref{fig:kotoIIdet}). 
Discovery of the decay with $5\sigma$ significance at the SM value of the branching ratio is achievable in $3\times 10^7$~s of KOTO~II running, 
with the collection of 35 signal events and 40 background events expected. 
An indication of new physics with a significance of 90\% is possible if the observed branching ratio differs by 40\% from the SM value.

The decay $K_L\to\pi^0 e^+e^-$ is also a sensitive prove to search for new physics sharing the same or similar short distance physics as
the $\klpionn$ decay. The $K_L\to\pi^0 e^+ e^-$ decay is also an
important input to study the flavor structure
as shown in Fig.~\ref{fig:LFU},
where the input from $K_L\to\pi^0 e^+ e^-$ constrains the parameters. This decay is the primary target in the second phase of the KOTO~II experiment,
in which the detector setup will be optimized with the addition of charged-particle tracking for the measurement of its branching ratio.

\paragraph{Detector development}
The KOTO-II Collaboration is developing numerous innovative detector technologies including a granular pre-shower counter to measure the incident angles of photons with an angular resolution of 1--3 degrees for photons with energies of several hundred MeV, calorimeter technologies with undoped CSI crystals and shashlyk counters, 
a lead and plastic scintillator sandwich counter incorporating $\mathrm{B}_4\mathrm{C}$ sheets to reduce the sensitivity to thermal neutrons, a fast crystal in-beam photon counter using ultra-fast lead
tungstate crystals with a scintillation decay constant of 640~ps, in-beam charged veto counters with thin fast timing silicon detectors, a plastic scintillator with quantum dots dispersed in UV-curable acrylic to obtain larger light yields and fast timing, a new 16 channel 500-MHz waveform digitizer.
\paragraph{European contributions}
Significant European contributions to KOTO~II are driving the development
of key aspects of the experiment. 

The idea of the measurement of the $K_L\to\pi^0 e^+e^-$ decay has been developed by the European members based on studies for the HIKE project.
The experience in the charged particle tracking in the NA62 experiment is a key to the design of the measurement. 
Straw tracker planes are under discussion
for the measurement of $K_L\to\pi^0 e^+e^-$ in the second phase. 
In-beam silicon charged veto counters are also under discussion.

A crystalline metal absorber to enhance the electromagnetic showers via  coherent interactions of photons incident along the crystal axis has been developed and is under consideration for use as the photon absorber in the KOTO~II beamline.
This will allow the thickness of the absorber to be reduced with no decrease in its effectiveness at reducing the photon flux in the beam, which will reduce $K_L$ losses and scattering in the absorber.

An in-beam photon veto counter with ultra-fast lead tungstate crystals has been developed for the HIKE project, and is under consideration for use in KOTO~II.
The fast timing will allow the use of a shorter veto timing window, thus 
reducing signal loss from accidental hits in the veto.

\paragraph{Timeline and Costs}
The KOTO experiment is under operation 
and is expected to reach its final goal by the end of the decade.
The start of the operation of the KOTO~II experiment is aimed in the early 2030s.
The detector cost is evaluated to be 12 MCHF~\footnote{The currency rate of 1 CHF=170 Yen is used.}.
An extension of the J-PARC Hadron Experimental Facility is needed to prepare the beamline and the experimental area of KOTO~II. The early realization of the extension is being discussed between members from KOTO~II and the other hadron-physics experiments hosted at the extension, and the management of the Institute of Particle and Nuclear Studies at KEK.
\section{Conclusion} 
KOTO will be the only dedicated rare kaon rare decay experiment in the world after 2026, and KOTO~II is the only currenlty planned future project. 
We envision KOTO~II as a global initiative for the experimental study of rare kaon decays, which has already received significant contributions from the European particle-physics community. In order to achieve this vision and successfully carry out the experimental program for the measurement of $\klpionn$ and $\klpioee$, we request support for European contributions for KOTO~II both for hardware development and physics exploitation.

\bibliographystyle{common/utphys}
\bibliography{common/koto2}
\end{document}